\documentclass[12pt,preprint]{aastex}

\usepackage{graphicx}
\usepackage{color}
\begin{document}

\title{The photospheric radiation model for the prompt emission of Gamma-ray Bursts: Interpreting four observed correlations}

\author{\sc Yi-Zhong Fan\altaffilmark{1,2}, Da-Ming Wei\altaffilmark{1,2}, Fu-Wen Zhang\altaffilmark{1,2,3} and Bin-Bin Zhang\altaffilmark{4}}
\altaffiltext{1}{Purple Mountain Observatory, Chinese Academy of Sciences, Nanjing 210008, China;}
\altaffiltext{2}{Key Laboratory of Dark Matter and Space Astronomy, Chinese Academy of Sciences, 210008, Nanjing, China;}
\altaffiltext{3}{College of Science, Guilin University of Technology, Guilin, Guangxi 541004, China;}
\altaffiltext{4}{Department of Astronomy and Astrophysics, Pennsylvania State University, University Park, PA 16802, USA.}
\email{yzfan@pmo.ac.cn,dmwei@pmo.ac.cn,fwzhang@pmo.ac.cn,bbzhang@psu.edu}
\begin{abstract}
We show that the experiential $E_{\rm p}-L$, $\Gamma-L$, $E_{\rm p}-\Gamma$ and ${\rm \bar{\eta}_\gamma}-E_{\rm p}$ correlations (where $L$ is the time-averaged luminosity of the prompt emission, $E_{\rm p}$ is the spectral peak energy, $\Gamma$ is the bulk Lorentz factor and $\bar{\eta}_\gamma$ is the emission efficiency of Gamma-ray bursts) are well consistent with the relations between the resembling parameters predicted in the photospheric radiation model of the prompt emission of Gamma-ray bursts. The time-resolved thermal radiation of GRB 090902B does follow the $E_{\rm p}-L$ and $\Gamma-L$ correlations. A reliable interpretation of the four correlations in alternative models is still lacking. These facts may point towards a photospheric origin of prompt emission of some Gamma-ray Bursts.
\end{abstract}


\setlength{\parindent}{.25in}

\section{Introduction}
In the past fifteen years, our understanding of gamma-ray bursts (GRBs) had been revolutionized. As usual, some aspects are understood better than others. For example, the detection of a bright supernova component
in the afterglow of some nearby long GRBs establishes
their collapsar origin and the late ($\sim 10^{4}$ s after the trigger of the burst) afterglow data support
the external forward shock model (Piran 2004; Zhang \&
M\'esz\'aros 2004).  Yet the physical origin of the prompt emission of GRBs is still not clear. The ``leading" internal shock model is found hard to explain some observational facts, motivating people to develop the internal magnetic energy dissipation models and the photosphere models \citep[see][for reviews]{Piran04,Zhang04}. It is rather hard to distinguish among these models reliably. It is widely speculated that the polarimetry of the prompt emission, for example, by POlarimeters for Energetic Transients (POET, Hill et al. 2008) and by POLAR (Orsi 2011),  may play key roles in the future. In this Letter we show that some empirical correlations of the prompt emission properties may have shed valuable light on the underlying physics and the photospheric model is favored.

\section{Interpreting the four observed correlations in the photospheric radiation model}

The tight correlation $E_{\rm p} \propto L^{0.5\pm0.1}$ was discovered by Wei \& Gao (2003, see Fig.6 therein) and then has been confirmed by many researches  \citep[e.g.,][]{Liang2004,Yonetoku2004,Ghirlanda2009,ZhangF2012}. Recently, a tight correlation $\Gamma \propto L^{0.3\pm0.002}$ was identified by \cite{Lv12} and a correlation $\Gamma \propto E_{\rm p}^{0.78\pm0.18}$ was suggested by \cite{Ghirlanda2012}. Very recently, \cite{Margutti2012} and \cite{Ber12} discovered a tight correlation $E_{\rm \gamma}/E_{\rm x} \propto E_{\rm p}^{0.66\pm0.16}$, where $E_\gamma$ is the isotropic-equivalent energy of the prompt emission and $E_{\rm x}$ is the total energy of the afterglow emission in X-ray band. In the forward shock afterglow model, $E_{\rm x}$ is proportional to $E_{\rm k}$, the kinetic energy of the outflow \citep{Piran04,Zhang04}. Therefore $E_{\rm \gamma}/E_{\rm x}(\propto E_{\rm \gamma}/E_{\rm k})$ is proportional to the GRB efficiency $\bar{\eta}_\gamma \equiv E_{\rm \gamma}/(E_{\rm \gamma}+E_{\rm k})$ as long as $E_{\gamma}$ is considerably smaller than $E_{\rm k}$. Hence one has $\bar{\eta}_\gamma \propto E_{\rm p}^{0.7}$. Some possible interpretations of the $E_{\rm p}-L$ correlation can be found in the literature \citep[e.g.,][]{Wei03,Rees2005,Ghirlanda2012}. In this Letter we aim to interpret all the above four correlations together \footnote{{ Two other highly relevant correlations are the $E_{\rm p}-E_{\rm \gamma,iso}$ correlation (Amati et al. 2002) as well as the $E_{\rm \gamma,iso}-\Gamma$ correlation (Liang et al. 2010), where $E_{\rm \gamma,iso}$ is the isotropic energy of the prompt $\gamma$-rays. Both of them are interpretable if one takes the duration of the bursts to be roughly a constant.}}. The starting point is the extensively discussed speculation that the prompt emission of Gamma-ray {bursts} is mainly from the photosphere which suffers significant modification and its spectrum is normally not thermal-like any longer \citep[e.g.,][]{Rees2005,Ioka2007,Beloborodov2010,Lazzati2011,Giannios2012}.

Firstly, we discuss {\emph{the simplest scenario}}, in which the (luminosity, spectral peak energy, efficiency) of the emission roughly resemble ($L_{\rm b}$, $T_{\rm b}$, $Y_{\rm b}$), where $(L_{\rm b},~T_{\rm b},~Y_{\rm b})$ {are} the (luminosity, temperature, efficiency) of the photospheric radiation, and $Y_{\rm b}$ and $L_{\rm b}$ are related to the total luminosity $L_0$ as $Y_{\rm b}=L_{\rm b}/L_0$. In such a scenario, if there are some valid correlations among $L_{\rm b}$, $T_{\rm b}$, $\Gamma$, and $Y_{\rm b}$, so are $L$, $E_{\rm p}$, $\Gamma$ and $\bar{\eta}_\gamma$.
For a relativistic baryonic fireball, the acceleration and the subsequent photospheric radiation  have been {initially} investigated by \cite{Piran93} and by \cite{Meszaros93}. Following these approaches, \cite{Fan2011} have {recently} derived the expressions of the initial radius of the {accelerated outflow} (i.e., $R_0$) and the final Lorentz factor of the outflow (i.e., $\Gamma$)
\begin{equation}
R_{0}\propto~L_{\rm b}^{1/2}Y_{\rm b}^{3/2}T_b^{-2},
\end{equation}
\begin{equation}
\Gamma \propto (Y_{\rm b}^{-1}-4/3)^{1/4}L_{\rm b}^{1/8}T_{\rm b}^{1/2},\label{eq:Gamma_ph}
\end{equation}
respectively. For $Y_{\rm b}\ll 1$ (actually even for $Y_{\rm b}=0.5$, the difference between $(Y_{\rm b}^{-1}-4/3)^{1/4}$ and $Y_{\rm b}^{-1/4}$ is only by a factor of $1.3$), eq.(\ref{eq:Gamma_ph})  reduces to the form obtained by \cite{Peer07}, i.e.,
\begin{equation}
\Gamma \propto Y_{\rm b}^{-1/4}L_{\rm b}^{1/8}T_{\rm b}^{1/2}.\label{eq:Gamma}
\end{equation}

As shown in \cite{Lv12}, for the outflow launched via the annihilation of neutrino pairs emitting from a hyper-accreting disk, the dimensionless entropy of the initial outflow is related to the total luminosity as $\eta \propto L_0^{\rm k}$ ({a} $k\sim 7/27$ is derived if the poorly understood collimation process is ignored \citep{Lv12}. In the following derivation we regard k as a ``free parameter"). The final Lorentz factor of the accelerated outflow is related to the initial dimensionless entropy as  $\Gamma \approx 4(1-4Y_{\rm b}/3)\eta/3$. As long as the thermal radiation is not extremely efficient (say, $Y_{\rm b}\leq 0.25$)\footnote{ The GRB efficiency of some bursts is quite high if one takes the energy injection model to account for the early shallowly decaying X-ray afterglow data. Such kind of models however are usually found to be unable to interpret the simultaneous optical afterglow data, as firstly pointed out by Fan \& Piran (2006). The modeling of the late ($t>10^{4}$ s) better-understood afterglow data suggests a typical GRB efficiency $\sim 10-20\%$ (e.g., Fan \& Piran 2006).}, approximately we have
\begin{equation}
\Gamma \propto L_{\rm b}^{\rm k}Y_{\rm b}^{\rm -k}.
\end{equation}
Combining eq.(1) with eq.(3), we have
\begin{equation}
\Gamma \propto L_{\rm b}^{1/4}R_0^{-1/4}Y_{\rm b}^{1/8}.
\end{equation}
Substituting this relation into eq.(4) we have
\begin{equation}
Y_{\rm b}\propto L_{\rm b}^{8k-2\over 1+8k}R_0^{2\over 1+8k}.
\end{equation}
Hence eq.(4) and eq.(1) give
\begin{equation}
\Gamma \propto L_{\rm b}^{3k \over 1+8k}R_0^{-{2k\over 1+8k}},
\end{equation}
and
\begin{equation}
T_{\rm b}
\propto L_{\rm b}^{{32k-5\over 4(1+8k)}}R_0^{1-4k \over 1+8k},
\end{equation}
respectively.
Finally we have
\begin{equation}
\Gamma \propto T_{\rm b}^{12k \over 32k-5}R_0^{-{2k\over (32k-5)}}.
\end{equation}
So far we have shown that some correlations should be present.

In the current scenario, $(E_{\rm p},~L,~\bar{\eta}_\gamma)$ largely resembles $(T_{\rm b},~L_{\rm b},~Y_{\rm b})$, respectively.
So if we take $k\sim 0.34$, the expected relations are
\begin{eqnarray}
\Gamma &\propto & E_{\rm p}^{0.7}R_0^{-0.11},~~\Gamma \propto L^{0.27}R_0^{-0.18},\nonumber\\
E_{\rm p} &\propto & L^{0.4}R_0^{-0.1},~~\bar{\eta}_{\gamma}\propto E_{\rm p}^{0.5}R_0^{0.5}, \label{eq:Main-results}
\end{eqnarray}
respectively, which are nicely in agreement with the four correlations summarized in the first paragraph of this section and the only {requirement} is that $R_0$ depends on $L$ insensitively.
{\it Interestingly}, the required $k\sim 0.34$ is close to that ($k\sim 7/27$) found in a simple analytical approach \citep{Lv12}. { Actually when adopting eq.(18) and eq.(16) of Fan \& Wei (2011), we have $\Gamma \approx 400 (L/10^{52}~{\rm erg~s^{-1}})^{1/4}(Y_{\rm b}/0.2)^{1/8}(R_0/10^{8}~{\rm cm})^{-1/4}$ and $E_{\rm p}\approx 260~{\rm keV}~(L/10^{52}~{\rm erg~s^{-1}})^{1/4}(Y_{\rm b}/0.2)^{3/8}(R_0/10^{8}~{\rm cm})^{-1/2}$, the coefficients are consistent with those reported in the literature, as long as $R_0$ is in order of $10^{8}$ cm. These facts together with the plots in Fig.1 illustrate that the correlations found in the literature (including the normalization) are indeed interpretable within the photosphere model.}

Secondly, we adopt the so-called {\emph{``generic" dissipative photospheric model}} developed by \citet{Giannios2012}, in which it is shown that at the radius $R_{\rm eq}$ (see eq.(5) therein), where radiation and electrons drop out of equilibrium, the spectral peak of the prompt emission forms \footnote{ The ``generic" dissipative photospheric model is different from the simplest photosphere model in two main aspects. One is that the electron-positron pairs delaying photosphere have been taken into account. The other is that the peak energy of the emerging spectrum traces the temperature of the outflow at $R_{\rm eq}$ (the optical depth is about tens, see eq.(6) therein) rather than that at the photospheric radius.} and the Lorentz factor can be expressed as (see eq.(9) therein)
\begin{equation}
\Gamma\propto E_{\rm p}^{3/5}\bar{\eta}_\gamma^{-1/5}L^{1/10}f_{\pm}^{1/5}(\eta/\Gamma)^{-1/5},
\label{eq:Gen-1}
\end{equation}
where $f_{\pm}$ is the number of electron+positron pairs per proton and is expected to be moderate. The acceleration calculation yields $R_{\rm eq} \propto \Gamma R_0 \bar{\eta}_\gamma^{-3/2}$  \citep[e.g.,][]{Piran93,Fan2011}, with which we have\footnote{ Numerically one gets
$\Gamma \approx 120(L/{10^{52}~{\rm erg~s^{-1}}})^{1/4}(\bar{\eta}_\gamma/0.2)^{1/4}(R_0/10^{8}~{\rm cm})^{-3/10}(f_\pm/5)^{1/5}(\eta/\Gamma)^{-1/5}$ and then $E_{\rm p} \approx  160~{\rm keV}~(L/{10^{52}~{\rm erg~s^{-1}}})^{1/4}(\bar{\eta}_\gamma/0.2)^{3/4}(R_0/10^{8}~{\rm cm})^{-1/2}$. These coefficients are comparable with that of the observed correlations as long as $R_0 \sim 10^{7}$ cm.}
\begin{equation}
\Gamma \propto L^{1/4}\bar{\eta}_\gamma^{1/4}R_0^{-3/10}f_\pm^{1/5}(\eta/\Gamma)^{-1/5}.
\label{eq:Gen-2}
\end{equation}
With the relation $\eta \propto L^{\rm k}\bar{\eta}_\gamma^{\rm -k}$, eq.(\ref{eq:Gen-1}) and eq.(\ref{eq:Gen-2}) give
\begin{eqnarray}
E_{\rm p} &\propto & L^{(10k-1)\over 6}\bar{\eta}_\gamma^{(1-5k)\over 3}f_\pm^{-1/3}(\eta/\Gamma)^{-4/3},\label{eq:Gen-3}\\
\bar{\eta}_\gamma & \propto & L^{(4k-1)\over 4k+1}R_0^{6\over 5(4k+1)}f_\pm^{-4\over 5(4k+1)}(\eta/\Gamma)^{-16\over 5(4k+1)},\label{eq:Gen-4}
\end{eqnarray}
respectively. Substituting eq.(\ref{eq:Gen-4}) into eq.(\ref{eq:Gen-2}) and eq.(\ref{eq:Gen-3}), we have
\begin{equation}
\Gamma \propto L^{2k \over 4k+1}R_0^{-{6k\over 5(1+4k)}}f_{\pm}^{4k\over 5(4k+1)}(\eta/\Gamma)^{-{4k+5 \over 5(4k+1)}},
\end{equation}
\begin{equation}
E_{\rm p}\propto L^{8k-1\over 2(4k+1)}R_0^{2(1-5k)\over 5(4k+1)}f_\pm^{-{3\over 5(4k+1)}}(\eta/\Gamma)^{-{12\over 5(4k+1)}},
\end{equation}
respectively.
As long as the radiation efficiency is not very efficient (say $\bar{\eta}_\gamma<0.25$), one can take $\eta/\Gamma \sim 1$ \citep{Piran93,Meszaros93}. For $k\sim 0.34$ we have \[\Gamma \propto L^{0.29},~E_{\rm p}\propto L^{0.37},~\Gamma\propto E_{\rm p}^{0.78},~\bar{\eta}_\gamma \propto E_{\rm p}^{0.4},\]which are roughly consistent with the correlations summarized at the beginning of this section.

{ Both long and short GRBs follow the $E_{\rm p}-L$ correlation \citep{Ghirlanda2009,ZhangF2012} and the $\bar{\eta}_\gamma-E_{\rm p}$ correlation \citep{Margutti2012,Ber12}. When taking the peak time of the GeV emission of the short GRB 090510 as the deceleration time of the forward shock, we found that the inferred bulk Lorentz factor also follows the $\Gamma-L$ correlation. Such facts suggest that the photospheric origin of the prompt emission may also apply to some short bursts.}

\begin{figure}
\includegraphics[width=85mm,angle=0]{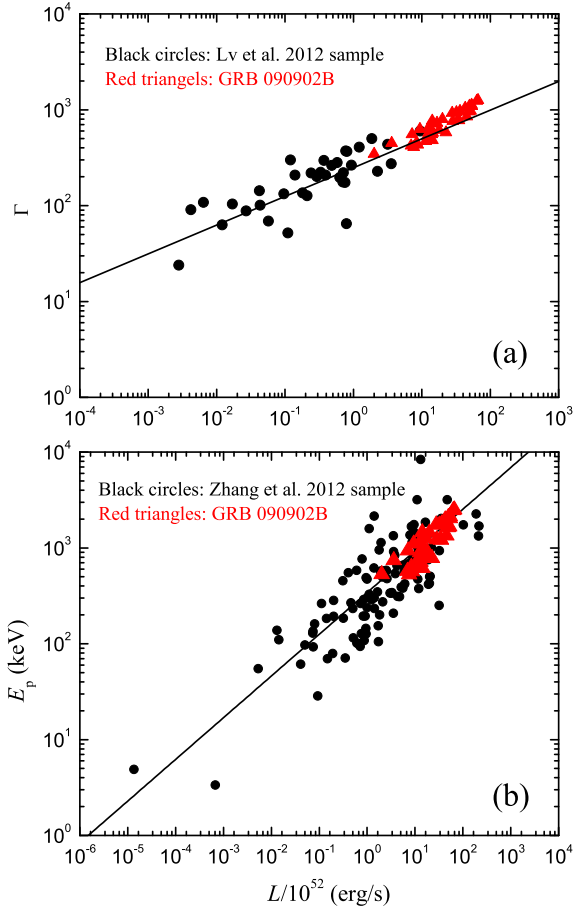}
\caption{(a) The $\Gamma-L$ diagram for the bursts discussed in \citet[][excluding those with a $\Gamma$ in dispute, for example GRB 090510 and GRB 090328A]{Lv12} and for the time-resolved thermal radiation of GRB 090902B. The solid line is the best fit $\Gamma\approx 249(L/10^{52}~{\rm erg~s^{-1}})^{0.3}$ obtained in \cite{Lv12}. (b) The $E_{\rm p}-L$ diagram for the bursts investigated in  \cite{ZhangF2012} and for the time-resolved thermal radiation of GRB 090902B (Please note that we have taken $E_{\rm p}=3.92(1+z)T_{\rm b,obs}$, where $T_{\rm b,obs}$ is the observed temperature). The solid line is the best fit $E_{\rm p}\approx 302~{\rm keV}~(L/10^{52}~{\rm erg~s^{-1}})^{0.4}$ found in \cite{ZhangF2012}.} \label{fig:Cartoon}
\end{figure}

\section{Discussion}
 Prominent thermal radiation components have been identified in GRB 090902B, a very bright burst at a redshift $z=1.822$ \citep{Abdo2009,Pandey2010,Ryde2010,ZhangBB2011,Liu2011,Barniol2011,Peer2012}. For example, \cite{ZhangBB2011} {divided} the whole data set of GRB 090902B into several time bins and {showed} that the spectrum in each bin can be nicely fitted by a thermal component plus a power-law spectral component. By applying the same technique, we redo the analysis using {\textit Fermi}/GBM data and the newest {\textit Fermi}/LAT PASS7 data. The thermal (blackbody) and non-thermal (power-law) spectral parameters and fluxes are derived in each time bin. Following \cite{Peer07} and \cite{Fan2011} and assuming a constant thermal radiation efficiency $\sim 20\%$, the bulk Lorentz factors of the outflow shells can be straightforwardly evaluated. We plot the inferred $\Gamma$ together with the simultaneous luminosity in the $\Gamma-L$ diagram presented by \cite{Lv12}. As shown in Fig.1(a) these two sets of data are in agreement with each other. For most bursts discussed in \cite{Lv12} the measurement of $\Gamma$ was based on the modeling of the afterglow light curve(s). The physics involved in such a kind of estimation is completely different from that for GRB 090902B. The agreement between these two sets of data thus not only supports our speculation of the photospheric origin of the prompt emission but also validates the robustness of both methods of evaluating $\Gamma$. In Fig.1(b) we plot the time-resolved spectral peak energy versus the simultaneous luminosity of GRB 090902B in the $E_{\rm p}-L$ diagram presented by \cite{ZhangF2012}. Again, a nice agreement between these two sets of data is present, in support of the photospheric origin of the prompt emission of some Gamma-ray bursts.

Finally, we'd like to point out that all these correlations have not been reasonably interpreted in either the internal shock models or the internal magnetic energy dissipation models (the outflow is magnetic). In the standard internal shock model, one has $E_{\rm p}\propto L^{1/2} \Gamma^{-2}$ \citep[e.g.,][]{Zhang02,Dai02,Fan05} then we expect no evident positive correlation between $E_{\rm p}$ and the luminosity after taking into account the correlation $\Gamma \propto L^{0.3}$, at odds with the data. It is also straightforward to show that the correlation $\Gamma \propto L^{0.3}$ predicts an extremely low internal shock efficiency unless the slow material shell has a width much widely than that of the fast shell (i.e., the duration of ejecting the slow shell is needed to be a factor of $\sim (\Gamma_{\rm f}/\Gamma_{\rm s})^{3.4}$ that of ejecting the fast shell, where $\Gamma_{\rm f}$ and $\Gamma_{\rm s}$ are the bulk Lorentz factor of the fast and slow shells, respectively). For a magnetic outflow, it is recognized in \cite{Lv12} that an interpretation of $\Gamma-L$ correlation is not available yet, let alone interpret the others. All these facts strongly favor the suggestion that the dominant component of the prompt emission of some GRBs may be tightly relevant to the photospheric radiation process, though much work on getting a spectrum nicely matching the data is still needed (Veres, Zhang \& Meszaros 2012, in preparation).

\acknowledgements
We thank B. Zhang, D. Giannios, R. Margutti and M. G. Bernardini for helpful communications and the referee for insightful comments. This work was supported in part by National Natural Science of China under grants 10973041, 10921063, 11073057 and 11163003, and by National Basic Research Program of China under grant 2009CB824800. YZF is also supported by the 100 Talents program of Chinese Academy of Sciences. {BBZ acknowledges the support from NASA SAO SV4-74018.}

\end{document}